\documentclass{article}
\usepackage[utf8]{inputenc} 
\usepackage[T1]{fontenc}
\usepackage{url}              % provides \url{...}
\usepackage{cite}             % improves presentation of citations

\usepackage[cmex10]{amsmath}  % Use the [cmex10] option to ensure complicance
\usepackage{mleftright}       % fix to wrong spacing of \left-,
\mleftright                   % \middle- \right-commands 

\usepackage{graphicx}         % provides \includegraphics{...} to
\usepackage{booktabs}         % fixes poor spacing in tables and
\usepackage{mathdots}
\usepackage[utf8]{inputenc} 
\usepackage[T1]{fontenc}
\usepackage{url}
\usepackage{ifthen}
\usepackage{cite}
\usepackage{algorithm}
\usepackage{algpseudocode}
\usepackage[cmex10]{amsmath} % Use the [cmex10] option to 
\usepackage{amssymb}
\usepackage{amsthm}
\usepackage{xcolor}

\usepackage{subcaption}
\usepackage{comment}
\usepackage{tabularx}
\usepackage{enumitem}
\setlist{nolistsep}
\usepackage{mathtools}
\usepackage{xcolor}
\usepackage{cancel}

\usepackage{tikz}
\usetikzlibrary {arrows.meta,automata,positioning} 

\usepackage{amsfonts}

\makeatletter
\newcommand{\multiline}[1]{%
  \begin{tabularx}{\dimexpr\linewidth-\ALG@thistlm}[t]{@{}X@{}}
    #1
  \end{tabularx}
}
\makeatother

\newtheorem{theorem}{Theorem}

\theoremstyle{definition}
\newtheorem{definition}{Definition}
\theoremstyle{remark}

\newtheorem{example}{Example}

\title{Algorithms for Computing the Free Distance of Convolutional Codes}
\author{Zita Abreu, Joachim Rosenthal, Michael Schaller}
\date{}

\begin{document}

\maketitle
\begin{abstract}
The free distance of a convolutional code is a reliable indicator of its performance. However its computation is not an easy task. In this paper,  we present some algorithms to compute the free distance with good efficiency that work for convolutional codes of all rates and over any field.
Furthermore we discuss why an algorithm which is claimed to be very efficient
is incorrect.
\end{abstract}

\section{Introduction}
Since real channels are inherently noisy, error correcting codes are necessary in all communication systems that handle digitally represented data.
Convolutional codes are one kind of error-correcting codes that is well-known for its suitability for sequential decoding with low delay.
This class of codes has been extensively researched, see \cite{johannesson1999fundamentals,bookchapter}.
%Currently, one of the key goals in the study of convolutional codes is to build convolutional codes of a specific rate and degree with a distance as large as possible.
The distance of a convolutional code indicates the code's robustness since it allows us to analyze its capacity to protect data against errors.
Codes with greater distance are preferable because they allow us to rectify more errors.
The free distance is one of the main types of distances for convolutional codes. This kind of distance is important for the decoding of a complete message. %Maximum Distance Separable Codes (MDS) are convolutional codes that have maximal free distance. Of all the convolutional codes with fixed rate and degree, these codes have the best error correcting performance. 

%In order to study and work with convolutional codes, efficient algorithms to calculate the free distance of convolutional codes are desired. 
In general, calculating the free distance of convolutional codes is not an easy task. Up to now, there are not many known algorithms that calculate the free distance. Actually, not much has been done lately to develop new algorithms or even improve old ones. %or simply to update these algorithms to the results that have been coming out over the years. 
Most algorithms in the literature are either too slow or use too much %computer memory 
computation and storage or are even wrong, see  e.g. \cite{david2000heapmod}, \cite{rouanne1989algorithm}, \cite{larsen1973comments}, \cite{bahl1972efficient} and \cite{cedervall1989fast}. Additionally, they are designed only over $\mathbb{F}_2$ and for certain rates. For instance, the authors of \cite{bahl1972efficient} and \cite{cedervall1989fast} made reference to the possibility of a generalization of the proposed algorithms, but they did not elaborate.

In this article we present some algorithms for computing the free distance of convolutional codes. In Section \ref{preliminaries} some introductory concepts are given. Section \ref{naive} provides a naive algorithm for calculating the free distance of convolutional codes.
In section \ref{heapmod} we prove that one of the existing algorithms for calculating the free distance published in \cite{david2000heapmod} is wrong.
In section \ref{sec:ofa}, we improve and generalize the algorithm provided in \cite{cedervall1989fast}, making it work for convolutional codes of all rates and over any field. Finally, in Section \ref{sec:newalg}, we present a novel algorithm for determining the free distance that also supports convolutional codes with all rates and degrees over any field. This algorithm combines the algorithm from Section \ref{sec:ofa} with the algorithm from article \cite{larsen1973comments}.

%Although our proposed modifications are conceptually straightforward, they significantly improve the algorithms for certain generator matrices.
%\textcolor{red}{Furthermore, we present a new algorithm for calculating the free distance of convolutional codes with better performance for certain generator matrices based on a combination of existing algorithms.} 

%Additionally, we discuss why the algorithm described in \cite{david2000heapmod} is wrong and why this comes down to the disproof of Miczo and Rudolph in \cite{MiczoR70} of a conjecture of Costello \cite{costello1969construction}. 

%\textcolor{red}{Finally, in addition to the algorithms we provide, we also make the implementations we made on Sagemath available on Github, see \url{https://github.com/uscpr/algorithms-for-computing-the-free-distance-of-convolutional-codes}.}

\section{Preliminaries}\label{preliminaries}

In this section, we provide a few key definitions and results for the following sections. For more details, see \cite{LinCostello} or \cite{bookchapter}.

First, denote by $\mathbb{F}_{q}[z]$ the ring of polynomials over the finite field with $q$ elements $\mathbb{F}_{q}$.

\begin{definition}
A \textbf{convolutional code} $\mathcal{C}$ of rate $k/n$ is a $\mathbb{F}_{q}[z]$-submodule of $\mathbb{F}_{q}[z]^n$ of rank $k$. A matrix $G(z)\in \mathbb{F}_{q}[z]^{k \times n}$ whose rows constitute a basis of $\mathcal{C}$ is known as a \textbf{generator matrix} for $\mathcal{C}$, i.e.:
%\vspace{-2mm}
{\begin{eqnarray} 
\mathcal{C} &\hspace{-1.5mm}
=\{{v(z) \in \mathbb{F}_{q}[z]^{n}: v(z) = u(z)G(z) \text{ with } u(z) \in \mathbb{F}_{q}[z]^{k}\}.}\nonumber
\end{eqnarray}}
\end{definition}

\begin{definition}
Let $G(z)=\sum_{i=0}^{\mu}G_i z^{i}\in\mathbb F_q[z]^{k\times n}$ with $G_{\mu} \neq 0$ and $k\leq n$. For each $i$, $1\leq i\leq k$, the $i$-th \textbf{row degree} $\nu_i$ of $G(z)$ is defined as the largest degree of any entry in row $i$ of $G(z)$, in particular $\mu=\max_{i=1,\hdots,k}\nu_i$. The \textbf{external degree} of $G(z)$ is the sum of the row degrees of $G(z)$. The \textbf{internal degree} of $G(z)$ is the maximal degree of the  $k\times k$ minors of $G(z)$ and it is the same for every generator matrix of the same code.
\end{definition}

\begin{definition}
A matrix $G(z)\in \mathbb F_q[z]^{k\times n}$ is said to be
\textbf{row reduced}\index{row reduced} if its internal and external degrees are equal. In this case, $G(z)$ is called a minimal generator matrix of the convolutional code it generates.
The \textbf{degree} $\delta$ of a code $\mathcal{C}$ is
the internal degree of a generator matrix of $\mathcal{C}$.
A convolutional code with rate $k/n$ and degree $\delta$ is known as a $(n,k,\delta)$ convolutional code.
\end{definition}

\begin{comment}
    \textcolor{red}{but there is something wrong with the definition of generic row degrees here, because if $k$ divides $\delta$, then all row degrees are the same but with the definition we have here this does not happen. I still need to check how to correct it but I better do other things first}
\end{comment}

\begin{definition}

The \textbf{free distance}\index{code!minimum distance}\index{distance!free} of a convolutional code
  $\mathcal{C}$ is given by
  \[d_{free}(\mathcal{C}):=\min_{v(z)\in\mathcal{C}}\left\{wt(v(z))\
    |\ v(z) \neq 0\right\},\]
     where $wt(v(z))$ is the Hamming weight of %a polynomial vector 
$v(z) = \sum_{t=0}^{\deg (v(z))}v_{t}z^{t} \in \mathbb{F}_{q}[z]^{n}$ that is defined as $wt(v(z)) = \sum^{\deg(v(z))}_{t=0} wt (v_t),$ where the weight $wt(v)$ of $v \in \mathbb{F}_{q}^{n}$ is the number of nonzero components of $v$.  
\end{definition}

%column distances
\begin{definition}
%For a polynomial vector ${v}(z)=\sum_{j\in\mathbb{N}_0}{v}_jz^j \in \mathbb{F}_q^n[z]$, the \textbf{j-th truncation}\index{truncation} of ${v}(z)$ is defined as
%\[
%{v}_{[0,j]}(z)={v}_0+{v}_1z+\cdots+{v}_jz^j.
%\]
For $j\in\mathbb N_0$, the \textbf{j-th column distance}\index{column distance} of a convolutional code $\mathcal{C}$ is defined as
\[
d_j^c(\mathcal{C}):=\min\left\{wt(v_0,\hdots,v_j)\ |\ {v}(z)\in\mathcal{C} \text{ and }{v}_0 \neq 0\right\}.
\]
\end{definition}

\begin{definition}\label{Def:Catas-and-Non-catas}
Let $\mathcal{C}$ be an $(n,k)$ convolutional code over $\mathbb{F}_q$. A full row rank matrix $H(z) \in \mathbb{F}_q[z]^{(n-k)\times n}$ satisfying
\[
\mathcal{C} = \ker H(z) = \{v(z) \in \mathbb{F}_q[z]^n \, : \, H(z)v(z)^\top = 0\}
\]
is called a \textbf{parity-check matrix} of $\mathcal{C}$. If this matrix exists, $\mathcal{C}$ is named \textbf{non-catastrophic}, otherwise it is named \textbf{catastrophic}.
\end{definition}

Moreover, from \cite{bookchapter} we know that a code is non-catastrophic if and only if $G(z)$ is left prime which is equivalent to $G(z)$ having finite output $u(z)G(z) \in \mathbb{F}_q[z]^n$ with $u(z) \in \mathbb{F}_q((z))^k$ implies that the input was already finite, i.e., $u(z) \in \mathbb{F}_q[z]^k$. 

\begin{definition}\label{revcode}
    Let $\mathcal{C}$ be a non-catastrophic convolutional code generated by  the row reduced matrix $G(z)$ with entries $g_{ij}(z)$.
    Let $\overline{g}_{ij}(z) = z^{\nu_i}g_{ij}(z^{-1})$ where $\nu_i$ is the $i$-th row degree.
    Then the code $\overline{\mathcal{C}}$ with generator matrix $\overline{G} = (\overline{g}_{ij})$ is called the \textbf{reverse code} of $\mathcal{C}$.
\end{definition}

In practice, instead of moving backward via the state-transition diagram of $\mathcal{C}$, one might go forward through the state-transition diagram of $\overline{\mathcal{C}}$.

The next theorem provides an upper bound for the free distance.

\begin{theorem}[$\hspace{-1mm}$ \cite{rosenthal1999maximum}]\label{th1} Let $\mathcal{C}$ be an $(n,k,\delta)$ convolutional code. Then, $d_{free}(\mathcal{C})\leq(n-k)\left(\left\lfloor
    \frac{\delta}{k}\right\rfloor+1\right)+\delta+1.$
This bound is called the \textbf{generalized Singleton bound}.
\end{theorem}
%An $(n,k,\delta)$ convolutional code with free distance equal to the generalized Singleton bound is called Maximum Distance Separable (MDS) convolutional code.

\section{Naive computation of the Free Distance}\label{naive}

This section provides a very simple method to calculate the free distance of a non-catastrophic convolutional code. The main idea behind this method is to find a non-trivial shortest path through the state-transition-diagram from the zero-state to the zero-state, with shortness assessed in terms of the weight of the output.

Consider $W^\ast$ to be an upper bound on the free distance. As example, assign $W^\ast$ to the generalized Singleton bound, see Theorem \ref{th1}.

Essentially, we will want to explore all paths until we either return to the zero-state and adjust the upper bound, or until the weight of the path exceeds $W^\ast$.

To ensure consistency across the algorithms provided here, we will describe this method by subtracting weights from the upper bound and from the states and checking if the residual weight is non-negative. Furthermore we move backwards through the state-transition-diagram.

Let us now describe the method as follows: We initialize an empty stack. We start at the zero-state and consider all possible extensions $S_E$ except for the ones that corresponds to the all zero input. 
If there are extensions $S_E$ from the zero-state to the zero-state which do not correspond to the all zero input (which can only happen if a row degree is $0$) we calculate the weights of the outputs corresponding to these extensions and set $W^\ast$ to the minimum of $W^\ast$ and these weights.
We do not put these extensions on the stack.
For all other extensions we set the weight $W_E$ of $S_E$ to be $W^\ast - w_E$ where $w_E$ is the weight of the output corresponding to the extension.
Then we store the extensions $S_E$ together with the weights $W_E$ in the stack if the weights are non-negative.
Next, we select a pair $(S, W)$ from the stack and look at all the extensions $(S_E, W_E)$.
The weight $W_E$ of an extension $S_E$ is now $W - w_E$. If the extension is the zero-state and if the weight $W_E > 0$, we change the upper bound $W^\ast$ to $W^\ast - W_E$ and we adjust the weight of the current state to $W - W_E$.
Moreover we adjust the stack, i.e., delete the states with weights less than $W_E$ and subtract $W_E$ from the weights of the remaining ones. If the extension is not the zero-state and if $W_E \geq 0$ we save $(S_E, W_E)$ on the stack.
We continue in the same way until the stack is empty.
Finally, we return $W^\ast$ which is the free distance.

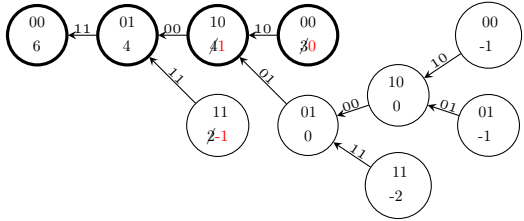
\begin{figure}[ht]
    \centering
    \tikzstyle{state}=[shape=circle, draw=black, align=center, scale=0.6]
    \tikzstyle{edge}=[->,>=stealth, scale=0.6]
    \tikzstyle{mainstate}=[state, very thick]
    \begin{tikzpicture}[scale=0.8]
        \node[mainstate] (s1_1) at (0,4) { $\begin{aligned} 00\\  6 \hspace{0.1cm} \end{aligned}$};
        \node[mainstate] (s1_2) at (1.5,4) {$\begin{aligned} 01 \\ 4 \hspace{0.1cm}\end{aligned}$}
            edge[edge] node[midway, above, yshift=-1mm, sloped] {\tiny{11}} (s1_1);
        \node[mainstate] (s1_3) at (3,4) {$\begin{aligned} 10 \\ \cancel{4}{\textcolor{red}{1}} \end{aligned}$}
            edge[edge]  node[midway, above, yshift=-1mm, sloped] {\tiny{00}}(s1_2);
        \node[state] (s2_3) at (3,2.5) {$\begin{aligned} \hspace{0.1mm} 11 \\ \text{{\cancel{2}{\textcolor{red}{-1}}}} \end{aligned}$}
            edge[edge] node[midway, above, yshift=-1mm, sloped] {\tiny{11}} (s1_2);
        \node[mainstate] (s1_4) at (4.5,4) {$\begin{aligned} 00 \\ \cancel{3}{\textcolor{red}{0}} \end{aligned}$}
            edge[edge] node[midway, above, yshift=-1mm, sloped] {\tiny{10}} (s1_3);
        \node[state] (s2_4) at (4.5,2.5) {$\begin{aligned} 01 \\ \text{\hspace{-0.1mm} 0 \hspace{-0.1mm}} \end{aligned}$}
            edge[edge] node[midway, above, yshift=-1mm, sloped] {\tiny{01}}(s1_3);
        \node[state] (s1_5) at (6,3) {$\begin{aligned} 10 \hspace{0.1cm}\\ \text{0 \hspace{0.05cm}} \end{aligned}$}
            edge[edge]  node[midway, above, yshift=-1mm, sloped] {\tiny{00}} (s2_4);
        \node[state] (s2_5) at (6,1.5) {$\begin{aligned} 11 \hspace{0.1cm}\\ \text{ {-2} \hspace{0.1cm}} \end{aligned}$}
            edge[edge] node[midway, above, yshift=-1mm,xshift=0.5mm, sloped] {\tiny{11}} (s2_4);
        \node[state] (s1_6) at (7.5,4) {$\begin{aligned} 00 \hspace{0.1cm}\\ \text{ {-1} \hspace{0.1cm}} \end{aligned}$}
            edge[edge] node[midway, above, yshift=-1mm, sloped] {\tiny{10}} (s1_5);
        \node[state] (s2_6) at (7.5,2.5) {$\begin{aligned} 01 \hspace{0.1cm}\\ \text{{-1} \hspace{0.05cm}} \end{aligned}$}
            edge[edge] node[midway, above, yshift=-1mm, xshift=0.5mm, sloped] {\tiny{01}} (s1_5);
    \end{tikzpicture}
    \caption{Naive computation of the free distance for a $(1,2,2)$ convolutional code with $G(z)=\begin{bmatrix}
    1 + z^2 & z^2
\end{bmatrix} \in \mathbb{F}_{2}[z]^{1 \times 2}$.}
    \label{fig1}
\end{figure}

To demonstrate the method, consider the following example in the Figure~\ref{fig1}.
\begin{example}
$W^\ast$ is initialized as $6$. While we move backwards in the state-transition-diagram, the only possible extension is the extension by $1$, i.e $S_1 = 01$.
In this state, $W_1 = W^\ast - w_1 = 6- wt(11)=6-2=4$. This is not the zero-state and $W_1 > 0$, so we add $(01,4)$ to the stack.
Since it is the only pair in the stack, we select it and look at all its extensions.
Extension $10$ will have weight $W-W_0=4-wt(00)=4-0=4$ and extension $11$ will similarly have weight $2$.
Again none of these states are the zero-state and $W_E > 0$ for both, so $(10,4)$ and $(11,2)$ are added to the stack and $(10, 4)$ will be on top of the stack.
Popping from the stack we get $S=10$.
Since, one of the extensions of $S$ is $S_0=00$ with $W_0=3$, we update the upper bound to $W^\ast=W^\ast-W_0=6-3=3$.
Additionally, we update $W$ to $W - W_0 = 4-3 = 1$.
Adjusting the stack, the state $11$ with weight $2$ is deleted from the stack.
We continue with the extension of $S = 10$ by $1$ to get $S_1 = 01$ with weight $0$.
Then we put it on the stack and pick it immediately again.
So $S = 01$ and $W = 0$.
We compute the extensions $S_0 = 10$ with weight $W_0 = 0$ and $S_1 = 11$ with $W_1 = -2$.
Only $(S_0, W_0)$ gets put on the stack.
Then we pick it from the stack.
Now both extensions have negative weight and the stack is empty, so we stop and return $W^\ast = 3$ as the free distance.
\end{example}

%Note that we could have stopped a little earlier, more precisely, we could have stopped when the weight was at most zero instead of negative.
%The way we described it is more interesting if one wants to compute the distance spectrum as in \cite{cedervall1989fast}.

Observe that, the algorithm will terminate since it will encounter repeated states for each path it can take in the tree.
As the code is non-catastrophic, the weight of the repeated state $S$ will be less than it was previously when encountered again.
Otherwise there would be a cycle of only zero output from $S$ to $S$, 
resulting in infinite input and finite output, which, by definition, does not occur if the code is non-catastrophic.
Obviously this argument provides only a rough upper bound for the number of steps that the algorithm performs.
In practice, it will terminate with considerably less steps than the bound. 

Overall, this naive computation of the free distance works over $\mathbb{F}_{q}$ and for convolutional codes of all rates, but it performs poorly. Therefore, we focused on developing better algorithms to calculate the free distance.

\section{The Heapmod algorithm is wrong}\label{heapmod}
In this section we study the algorithm presented in \cite{david2000heapmod}, called Heapmod algorithm and claimed to be better than the one published in \cite{cedervall1989fast}, and we demonstrate that it is incorrect. To be more specific, we rewrite the stop condition of the algorithm to provide a clearer understanding of its incorrectness. With our version of the stop condition, it is evident that the incorrectness is related to an old conjecture by Costello \cite{costello1969construction} and a class of counterexamples presented in \cite{MiczoR70} by Miczo and Rudolph.

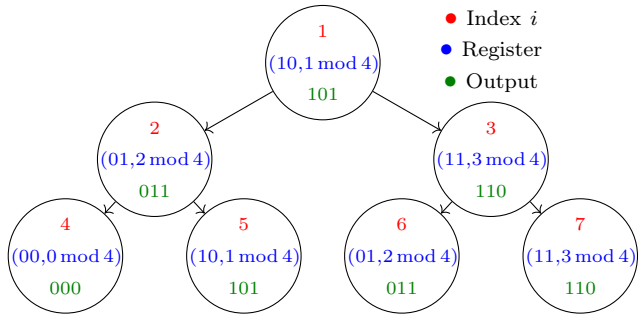
\begin{figure}[ht]
  \centering
  \begin{tikzpicture}[node distance=1cm, scale=0.1]
    \node[circle, draw, align=center, text width=0.655cm] (1) {
      \textcolor{red}{$\scriptstyle{1}$}\\
      $\textcolor{blue}{\hspace{-0.4cm}\scriptstyle{(10,1\hspace{-0.15cm}\mod \hspace{-0.05cm}4})}$ \\ 
      \textcolor[rgb]{0,0.5,0}{$\scriptstyle{101}$}
    };
    \node[circle, draw, align=center, below left= 0.2cm and 1.15cm of 1, text width=0.655cm] (2) {
      \textcolor{red}{$\scriptstyle{2}$}\\
      $\textcolor{blue}{\hspace{-0.4cm}\scriptstyle{(01,2\hspace{-0.15cm}\mod \hspace{-0.05cm}4})}$ \\ 
      \textcolor[rgb]{0,0.5,0}{$\scriptstyle{011}$}
    };
    \node[circle, draw, align=center, below right= 0.2cm and 1.15cm of 1, text width=0.655cm] (3) {
      \textcolor{red}{$\scriptstyle{3}$}\\
      $\textcolor{blue}{\hspace{-0.4cm}\scriptstyle{(11,3\hspace{-0.15cm}\mod \hspace{-0.05cm}4})}$ \\ 
      \textcolor[rgb]{0,0.5,0}{$\scriptstyle{110}$}
    };
    \node[circle, draw, align=center, below left=0.2cm and 0.1cm of 2, text width=0.655cm] (4) {
      \textcolor{red}{$\scriptstyle{4}$}\\
      $\textcolor{blue}{\hspace{-0.4cm}\scriptstyle{(00,0 \hspace{-0.15cm}\mod \hspace{-0.05cm}4})}$ \\ 
      \textcolor[rgb]{0,0.5,0}{$\scriptstyle{000}$}
    };
    \node[circle, draw, align=center, below right=0.2cm and 0.1cm of 2, text width=0.655cm] (5) {
      \textcolor{red}{$\scriptstyle{5}$}\\
      $\textcolor{blue}{\hspace{-0.4cm}\scriptstyle{(10,1 \hspace{-0.15cm}\mod \hspace{-0.05cm}4})}$ \\ 
      \textcolor[rgb]{0,0.5,0}{$\scriptstyle{101}$}
    };
    \node[circle, draw, align=center, below left=0.2cm and 0.1cm of 3, text width=0.655cm] (6) {
      \textcolor{red}{$\scriptstyle{6}$}\\
      $\textcolor{blue}{\hspace{-0.4cm}\scriptstyle{(01,2 \hspace{-0.15cm}\mod \hspace{-0.05cm}4})}$ \\ 
      \textcolor[rgb]{0,0.5,0}{$\scriptstyle{011}$}
    };
    \node[circle, draw, align=center, below right=0.2cm and 0.1cm of 3, text width=0.655cm] (7) {
      \textcolor{red}{$\scriptstyle{7}$}\\
      $\textcolor{blue}{\hspace{-0.4cm}\scriptstyle{(11,3 \hspace{-0.15cm}\mod \hspace{-0.05cm}4})}$ \\ 
      \textcolor[rgb]{0,0.5,0}{$\scriptstyle{110}$}
    };
    \node[above=0.01mm of 3, text width=4cm, align=center] {
      \textcolor{red}{\textbullet} {\footnotesize{Index $i$}} \\
      \textcolor{blue}{\textbullet} {\footnotesize{Register}}\\
      \textcolor[rgb]{0,0.5,0}{\textbullet} {\footnotesize{Output}}
    };
    \draw[->] (1) -- (2);
    \draw[->] (1) -- (3);
    \draw[->] (2) -- (4);
    \draw[->] (2) -- (5);
    \draw[->] (3) -- (6);
    \draw[->] (3) -- (7);
  \end{tikzpicture}
  \caption{Illustration of the Heapmod algorithm.}
  \label{fig2}
\end{figure}

The authors of the Heapmod algorithm construct the code tree in a slightly different manner than usual. Namely, instead of having the state on the nodes and the outputs on the edges they have the state together with the input on the node and there are no outputs on the edges. A decimal representation of registers is used, where the registers consist of input and state. To clarify, see figure \ref{fig2} to check how the first part of the tree looks like.
Note that the entire tree is not drawn.

In this section we let $\mathcal{C}$ be a $(n, 1, M)$ convolutional code.
We state the algorithm below, using the same notation as in \cite{david2000heapmod}.
\begin{enumerate}
    \item Calculate the weight $w(O_i)$ of the output $O_i$ corresponding to register $i$ for $i \in \lbrace 1, \ldots, 2^{M+1}-1 \rbrace$.
    \item Compute
    \[
        l_j = (2j + 1) \cdot 2^M
    \]
    for $j \in \lbrace 0, \ldots, 2^{M+1} - 1 \rbrace$.
    \item For each $j$ form a path from $l_j$ to the root of the tree and sum the weights of the corresponding outputs.
    The minimum over $j$ of these sums is the free distance.
\end{enumerate}

Taking into account the tree with $M=1$ whose first part is represented in Figure \ref{fig2}, the indices $l_j$ would be $2, 6, 10, 14$.
Note that coming to $l_j$ corresponds to a return to the zero-state.
This is easy to see, since $l_0 = 2^M$ and $l_{j+1} = (2(j+1) + 1) \cdot 2^M = (2j + 1 + 2) \cdot 2^M = l_j + 2^{M+1}$.
In this way, adding $2^{M+1}$ gives us exactly the next occurrence of the same register.
Therefore, let us reformulate the algorithm in the following way: It considers the $2^{M+1}$ shortest (in terms of steps through the state-transition-diagram) paths from the zero-state to the zero-state.

A simple counterexample for the correctness of the Heapmod algorithm is the non-catastrophic code generated by
\[
    G(z) = \begin{pmatrix}
        z^6 + z^4 + 1, & z^6 + z^5 + z^4 + z^3 + z + 1
    \end{pmatrix}.
\]
For this example, the Heapmod algorithm returns a free distance of $9$ instead of $8$ (which was the correct answer).
To be explicit $wt((1 + z^4 + z^6 + z^8) G(z)) = wt(\begin{pmatrix}
        z^{14} + 1, & z^{14} + z^{13} + z^9 + z^3 + z + 1
    \end{pmatrix})= 8$.
Similar examples can easily be obtained.

The underlying reason is the following: For each $j$ the length from $l_j$ to the root
 of the tree is at most 
\[
    \log_2(l_j) \leq \log_2(2 \cdot 2^{M+1} \cdot2^M) = 2M + 2.
\]
That means that the algorithm explores only inputs up to length $2M + 2 - M = M + 2$ since we also need $M$ zero inputs at the end to arrive at the registers corresponding to the $l_j$.
Note that in the example above the input length is $9$ since the degree of the input polynomial is $8$. Furthermore, since the input length is 9, it is evident that the algorithm does not explore everything because $M = 6$ and the algorithm only explores inputs up to length $M + 2 = 8$.

It was a conjecture of Costello in \cite{costello1969construction} that inputs of length $2M$ would be enough to determine the free distance.
In the same paper he also proved an upper bound quadratic in $M$ for the input length for systematic codes.
Miczo and Rudolph proved in \cite{MiczoR70} that no linear upper bound on the input length is enough to get the free distance, which also gives another proof of the incorrectness of the Heapmod algorithm.

\section{The optimized FAST algorithm}\label{sec:ofa}

In this section we will introduce an optimized version of the FAST algorithm, a stack algorithm which computes the free distance of convolutional codes, presented in \cite{cedervall1989fast}. More precisely, we present a generalized version that works for arbitrary convolutional codes of all rates and over any field, we explain the modifications we made and we address their impact in relation to the efficiency of the algorithm.

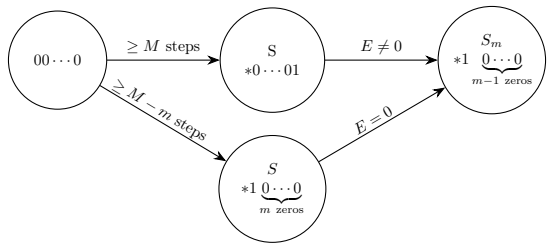
\begin{figure}[ht]
  \centering
  \begin{tikzpicture}[node distance=2.5cm, scale=0.6, transform shape]
    \node[circle, draw, align=center, text width=1.9cm] (1) {$0 0 \cdots 0$};
    \node[circle, draw, align=center, right=of 1, text width=1.9cm] (2) {S\\$*0\cdots 01$};
    \node[circle, draw, align=center, below=0.5cm of 2, text width=1.7cm] (3) {$S$\\$*1\underbrace{0\cdots0}_{m \mathrm{\ zeros}}$};
    \node[circle, draw, align=center, right=of 2, text width=1.7cm] (4) {$S_m$\\$*1\underbrace{0\cdots0}_{m-1 \mathrm{\ zeros}}$};
    \draw[->,>=Stealth] (1) -- (2) node[midway, above] {$\geq M$ steps};
    \draw[->,>=Stealth] (1) -- (3) node[midway, left, xshift=1.1cm, yshift = 2mm, sloped] {$\geq M-m$ steps};
    \draw[->,>=Stealth] (2) -- (4) node[midway, above] {$E \neq 0$};
    \draw[->,>=Stealth] (3) -- (4) node[midway, right,xshift=-0.7cm, yshift = 2mm, sloped] {$E=0$};
  \end{tikzpicture}
  \caption{Idea behind the (optimized) Fast algorithm for a convolutional code of rate $1/n$ over $\mathbb{F}_2$.}
  \label{fig3}
\end{figure}

In the Fast algorithm, the last non-zero entry in a state $S$ tells us the minimum number of steps $u$ we must go in the backward direction to reach the zero-state.
Additionally, moving forward from the zero-state to $S$ requires at least $u$ inputs, which tells us that the weight of this path is at least the $u-1$-st column distance $d_{u-1}$. This means that if the weight of a state $S$ is less than $d_{u-1}$ we do not have to extend this state, see Figure \ref{fig3}.

%The Fast algorithm will then need to pre-calculate the column distances. 

\begin{figure}[ht]
  \centering
\includegraphics[scale=0.4]{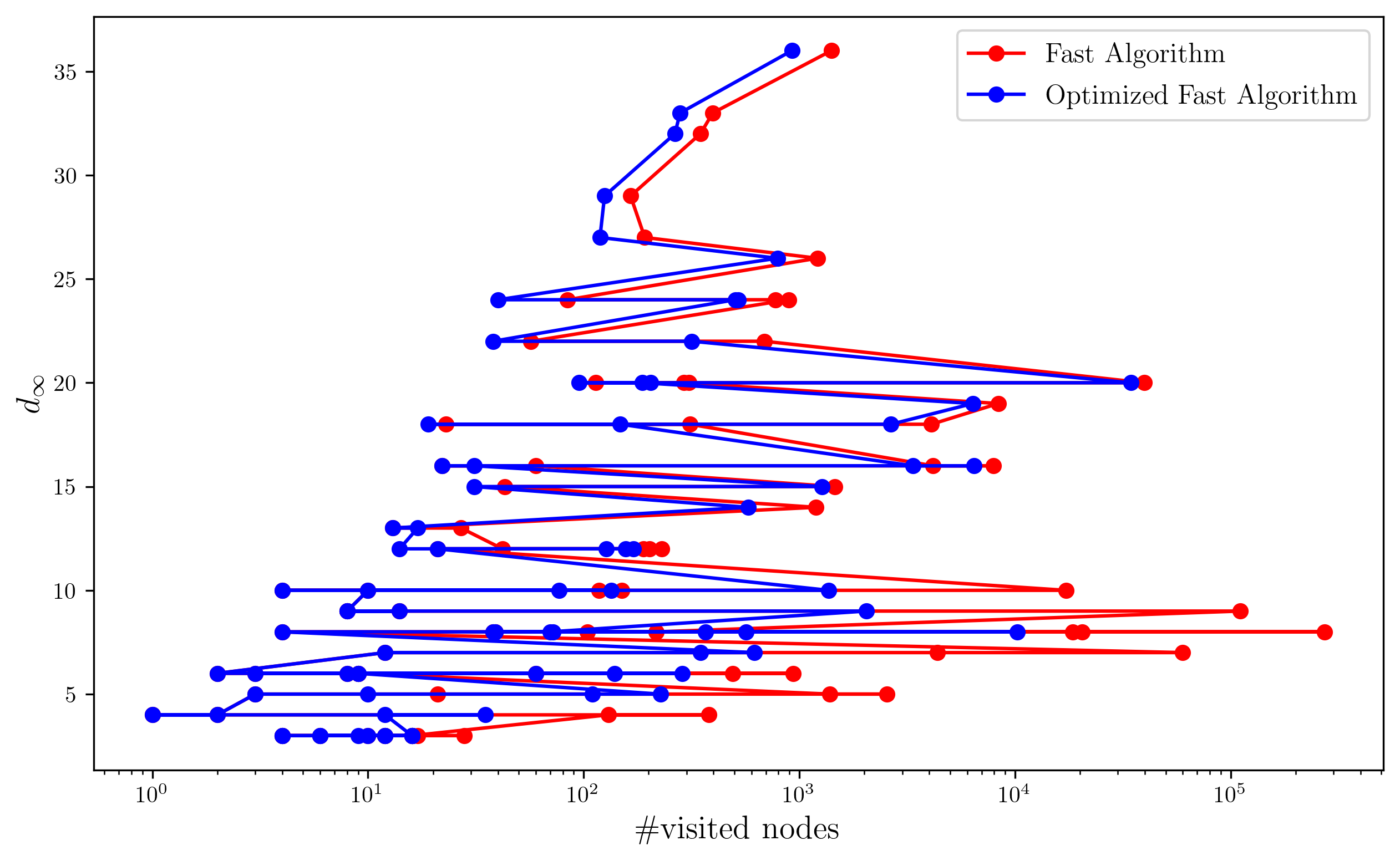}
  \caption{Efficiencies of the FAST algorithm and the optimized FAST algorithm.}
  \label{graphic1}
\end{figure}

\begin{figure}[ht]
  \centering
\includegraphics[scale=0.4]{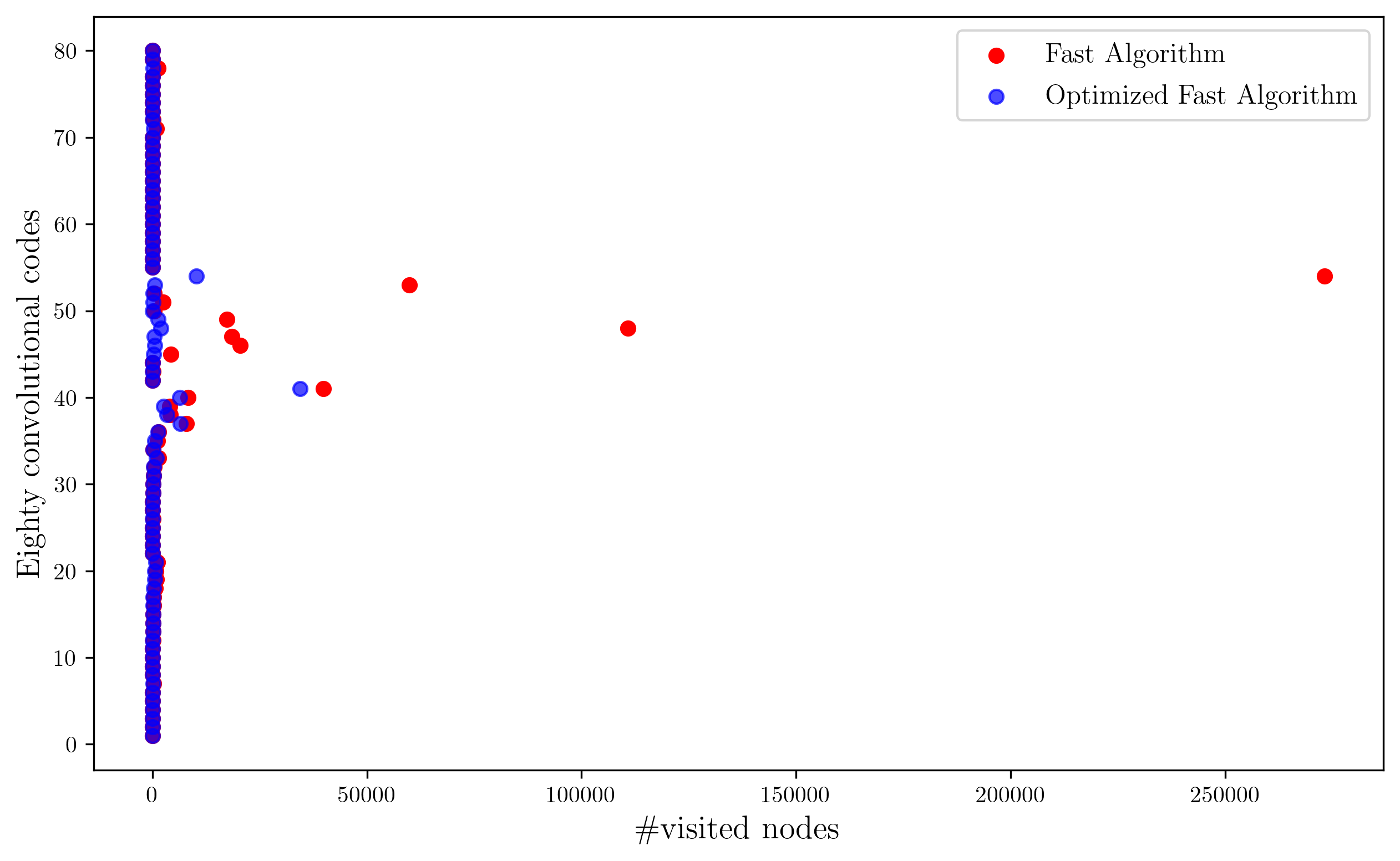}
  \caption{Another perspective of Figure \ref{graphic1}.}
  \label{graphic2}
\end{figure}

The optimized Fast algorithm aims to avoid unnecessary state extensions in the code tree that are not considered in the Fast algorithm. If one calculates the column distances $[d_0, \ldots, d_M]$ using brute force one can improve this by storing the actual distance of the state to the zero-state.
Here distance refers to the path with the fewest steps, which is not always the path with lowest weight.
This costs more memory, but can make the calculation considerably faster.
See Figure \ref{graphic1} and \ref{graphic2}, which illustrates the difference in efficiency between the Fast algorithm and the optimized Fast algorithm for eighty arbitrarily chosen convolutional codes from the literature with different rates and degrees and over distinct fields.
Figure \ref{graphic1} correlates the number of new nodes visited to the free distance of the chosen convolutional codes, whereas Figure \ref{graphic2} shows the number of nodes visited for each convolutional code separately. Both show a significant reduction in the number of nodes we need to visit using the optimized Fast algorithm.

Let us first introduce the notation for presenting the optimized Fast algorithm (Algorithm~\ref{alg:cap_fast}). The integer $m(S)_i$ represents the number of zeros to the right of the rightmost non-zero entry in row $i$ of state $S$. The $d(S)$ is the distance of a state to zero where we only consider paths of length at most $M+1$, where $M$ is the memory. This distance will be precomputed when we compute the column distances. Finally, the $j$-th column distance is denoted by $d_j$.

\begin{algorithm}
\caption{OPTIMIZED FAST ALGORITHM}\label{alg:cap_fast}
Given a generator matrix $G$ of a convolutional code $\mathcal{C}$ with row degrees $\nu_i$ and $W^*$ an upper bound of the free distance, we search the code tree of $\mathcal{C}$ to determine the free distance.
\begin{algorithmic}
\State Calculate $d_{j}$ with $j=0, \ldots, M$ and the distance $d(S)$ of the state $S$ to the zero-state.
\State Calculate all extensions $S$ of the zero-state corresponding to non-zero input together with their weights $W$.
If $S$ ends in the zero-state set $W^\ast = \min(W^\ast, W)$ and do not put the extension on the stack.
For the states $S$ which do not end in the zero-state put the tuples $(S, W)$ on the stack if $W \geq 0$.
\While {stack is nonempty}
    \State Take $(S, W)$ from the stack.
    \State Compute all extensions $S_E$ of $S$ and their weights $W_E$.
    \State Go through all of them.
    \If {$S_E$ is the zero-state and $W_E \geq 0$}
    \State \multiline{%
    Set $W^* = W^* - W_E$, adjust the weights of the extensions and of the states on the stack and continue.
    }
    \Else
    \State $\sigma = \max(\nu_i - m(S_E)_i)$
        \If 
        {\hspace{-0.1cm}
            (($W_E \hspace{-0.9mm} < \hspace{-0.9mm}  d(S_E)$ and $W_E \hspace{-0.9mm} < \hspace{-0.9mm} d_M$) or 
            $ W_E \hspace{-0.9mm} < \hspace{-0.9mm}  d_{\sigma - 1}$)
        \hspace{-0.1cm}}
        \State Discard this extension.
        \EndIf
    \EndIf
    \State \multiline{
        Pick one of the extensions that are left and set it to $S$ and its weight to $W$.
    }
    \State \multiline{
        Put the other extensions that are left on the stack.
    }
\EndWhile
\State Return $W^{\ast}$
\end{algorithmic}
\end{algorithm}

For a better understanding of the algorithm, consider that the variable $\sigma$ represents the minimum number of steps required to return to the zero-state. Note that this variable is used in the second part of the second if in order to discard the extension.

Regarding the first part of the second if, the condition $W_E < d(S_E)$ implies that we cannot return to the zero-state from $S_E$ in less than $M+1$ steps.
To get to the zero-state, we need to take at least $M + 1$ steps.
If $W_E < d_M$, we also cannot return to the zero-state in $M+1$ or more steps.

Let us conclude with some observations on the algorithm's implementation.
We recommend to pick the extension by zero if it is one of the remaining extensions.
Returning to the zero-state can lead to an adjustment of the upper bound, which can significantly reduce the runtime.
It may even make sense to employ more advanced heuristics to swiftly return to the zero-state and, hopefully, adjust the upper bound.

\section{A new algorithm}\label{sec:newalg}

In this section, we propose a new algorithm for computing the free distance of convolutional codes that is based on a combination of two algorithms, namely the optimized FAST algorithm from Section \ref{sec:ofa} and the algorithm proposed by Larsen in \cite{larsen1973comments}. The last one is a bidirectional algorithm that extends paths forward and backward simultaneously. Our algorithm is also bidirectional and it works for arbitrary convolutional codes of all rates and over any field.

This new algorithm also uses a state-transition diagram and keeps the notation used of Algorithm \ref{alg:cap_fast}. However, now when a path reaches a state, we stored it in an array together with the information about the path's type (forward or backward) and the weight. If there are many paths to a state, the lowest weight is stored. 

In this algorithm we will also need to calculate the weight by starting from the zero-state with weight $0$ and increasing it.
We denote this weight by $W_L$.
Note that, for the weight $W$ which we get by starting from the zero-state with weight $W^\ast$ and decreasing it we get $W + W_L = W^\ast$.

The following new notation must be fixed
\begin{itemize}
%\item $S$ - Terminal state of path.
%\item \textcolor{red}{$W_L$ - Minimum Hamming weight of paths to $S$ known at the moment. ???}
\item $T$ - Type of state $S$. The type consists of two parts, $T1$ and $T2$. $T1$ indicates whether the state $S$ is $D$(ead) or non$-D$ and $T2$ is the type, $F$(orward) or $B$(ackward), of the minimum-weight path first found to $S$. Thus we have four possible types: $F$, $B$, $DF$, and $DB$.
\end{itemize}

Note that if a variable depends on $T$ we use the same notation as in Algorithm \ref{alg:cap_fast} in case $T2 = B$.
If $T2 = F$, we use the same variables as in the fast algorithm but for the reverse code generated by the generator matrix $\overline{G}$.
By $G(T)$ we denote $G$ if $T = B$ and $\overline{G}$ if $T = F$.
Similarly $\nu_i(T)$ is the $i$-th row degree of $G$ if $T = B$ and the $i$-th row degree of $\overline{G}$ if $T = F$.

The variable $W^\ast$ remains the current upper bound of the free distance and the new algorithm for a rate$-k/n$ convolutional code is stated as Algorithm \ref{alg:cap_new}. In the algorithm where there is \textbf{continue} we skip all further steps in the for-loop and proceed in the for-loop with the next extension $E$.

\begin{algorithm}
\caption{NEW ALGORITHM}\label{alg:cap_new}
Given a generator matrix $G$ of a convolutional code $\mathcal{C}$ with row degrees $\nu_i$ and $W^*$ an upper bound of the free distance, we search the code tree of $\mathcal{C}$ to determine the free distance.

\begin{algorithmic}
\State 1) Start at the zero-state and calculate all forward and backward extensions $S$ and their weights $W_L$.
If some of these extensions have non-zero input and give the zero-state adjust the upper bound $W^\ast$.
We do not save these extensions.
We store all other extensions with the following exception:
If a $B$-extension is equal to some $F$-extension save only the one with lower weight $W_L$ and adjust the upper bound if the sum of the weights is less than $W^\ast$.
Together with a state $S$ we also store its weight $W_L$ and type $T$.
\While{not {}\textit{Done}}
\State \multiline{
2) Search through the storage array for the lowest weight non$-D$ state $S$. It has weight $W_L$ and is of type $T$.
}
\For{extensions $E$}
    \State \multiline{
    3) Determine terminal state $S_E$ and weight $W_{E, L}$ of the $E$-extension of $S$.
    } 
    \State 4) \multiline{If $S_E$ is the zero-state set $W^\ast = \min(W_{E, L}, W^\ast)$ and \textbf{continue}.}
    \State \multiline{
    5) $\sigma(T) = \max(\nu_i(T) - m(S_E, T)_i)$.
    If $W_{E, L} >\frac{W^\ast + n - 1}{2}$ or ($W_{E} \hspace{-0.9mm} < \hspace{-0.9mm}  d(S_E, T)$ and $W_{E} \hspace{-0.9mm} < \hspace{-0.9mm} d_M(T)$) or 
            $ W_{E, L} \hspace{-0.9mm} < \hspace{-0.9mm}  d_{\sigma - 1}(T)$ \textbf{continue}.
    }
    \State \multiline{
    6) Check through the array for $S_k = S_E$. If no such $S_k$ is found, store $S_E, W_{E, L}, T$ in the array and \textbf{continue}.
    } 
    \State \multiline{
    7) The type and weight of $S_k$ are $T_k$ and $W_{k, L}$. If ${T2}_k = T2$, go to 11).
    }
    \State 8) Set $W^\ast = \min(W^\ast, W_{E, L} + W_{k, L})$.
    \State 9) If $W_{E, L} \geq W_{k, L}$, \textbf{continue}.
    \State 10) Set $W_{k, L} = W_{E,L}$, ${T2}_k = T2$ and \textbf{continue}.
    \State 11) If $T1_k = D$, \textbf{continue}.
    \State 12) Set $W_{k, L} = \min (W_{k, L}, W_{E, L})$.
\EndFor
\State 13) Set $T1 = D$.
\State 14) Set $\textit{Done} =  (2W_L \geq W^\ast$ or all states are $D)$.
\EndWhile
\State \Return{$d_{free} = W^{\ast}$.}
\end{algorithmic}
\end{algorithm}

%The algorithm is based on Larsen's algorithm in \cite{larsen1973comments}.
%See the paper by Larsen to see what most of the steps do.
%Actually the proof of correctness there is incomplete as we will show in the appendix, but this can be remedied.

Regarding paper \cite{larsen1973comments}, it is interesting to note that a proof of correctness of the algorithm proposed there is incomplete. In fact, we demonstrate this in the Appendix and we explain why this can be remedied.

The main improvements of the new algorithm compared to the %original 
Algorithm in \cite{larsen1973comments} are that this algorithm works for all rates, degrees and over any field and that the algorithm visits fewer states.

In short, regarding the algorithm, step 4 is required in order to generalize it. More precisely, without this step the algorithm is wrong if $k > 1$. Additionally, with this step we also circumvent the  issues in the proof of the algorithm proposed by Larsen, which we elaborate on in the Appendix. Step 5, on the other hand, allows us similarly as in the optimized Fast algorithm to skip certain states, namely avoiding unnecessary state extensions in the code tree.

\begin{figure}[ht]
  \centering
\includegraphics[scale=0.4]{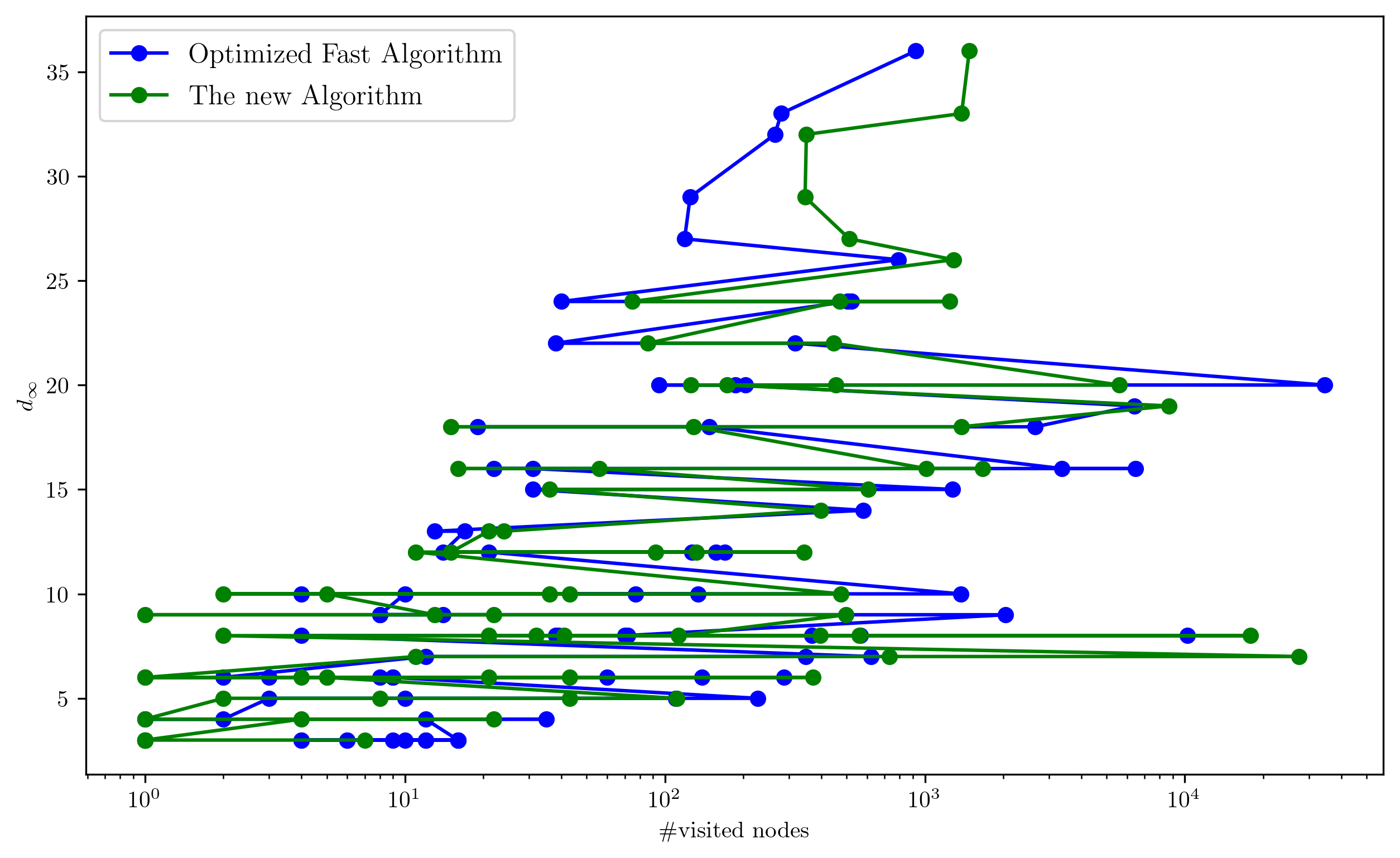}
  \caption{Efficiencies of the Optimized Fast algorithm and the new algorithm.}
  \label{final}
\end{figure}

Finally, Figure \ref{final} shows how the new algorithm compares to
%combination of all of this speeds up the computation when compared with 
the Optimized Fast Algorithm for the same eighty convolutional codes as before.
Note that this comparison is not entirely fair since the new algorithm has to search through an array to visit the nodes.

\section{Conclusion and Future Work}
Some algorithms to compute the free distance of convolutional codes were presented. Depending on the convolutional code the bottleneck of these algorithms might either be the number of states visited or the precomputation of the column distances. Therefore an efficient method for computing the $j$-th column distance is an interesting topic for further research. 

Additionally, for constructing convolutional codes with good free distance it will be important to have algebraic criteria. The state-transition diagram can be represented algebraically by linear system representations, see \cite{liro21}. It would be therefore interesting to review the new algorithm presented here in the algebraic systems representation.

To see the implementations of the algorithms mentioned here, visit \url{https://github.com/uscpr/algorithms-for-computing-the-free-distance-of-convolutional-codes}.

\appendix
\section{Comments on Larsen's Algorithm}

In this section, we will review the algorithm presented by Larsen in \cite{larsen1973comments}. We will demonstrate that the proof of the correctness of the mentioned algorithm is incomplete. However, despite this, the algorithm works. Therefore, we generalize it and use it as a reference for the construction of the new Algorithm %\ref{alg:cap_new}.
presented in Section VI.

In general, the proof the correctness of the algorithm is incomplete because it only implicitly takes the structure of the convolutional code into account. In fact,  it is easy to construct a state-transition diagram that looks like the state-transition diagram of a convolutional code but on which the algorithm fails if we are allowed to choose the weights on the edges as we wish.
Consider the following example.

\begin{example}\label{example}
Take the following state-transition diagram with the respective weights on the edges which has the same graph as a state-transition diagram for a convolutional code of memory $2$, see Figure \ref{state-transitiondiagram}.
    \begin{figure}[ht]
    \centering
    \begin{tikzpicture}[->,>={Stealth[round]},shorten >=1pt,
                    auto,node distance=2cm,on grid,semithick,
                    inner sep=1pt,bend angle=45,scale=0.6]
  \node[state,scale=1.1] (01)                    {$01$};
  \node[state,scale=1.1]         (11) [above right=of 01] {$11$};
  \node[state,scale=1.1]         (00) [below right=of 01] {$00$};
  \node[state,scale=1.1]         (10) [below right=of 11] {$10$};

  \path [every node/.style={font=\footnotesize}]
        (01)edge   [bend left,scale=0.7]            node[swap,yshift = -1.5mm] {{20}} (10)
            edge   [bend right]            node [swap]{{10}} (00)
        (11) edge [loop above,scale=0.7] node {{20}} (11)
            edge  [bend right,scale=0.7]             node [swap]{{0}} (01)
        (10) edge   [bend right,scale=0.7]            node [swap]{{1}} (11)
             edge   [bend left]            node [swap, yshift=1.5mm]{{2}} (01)
        (00) edge [loop below,scale=0.7] node {{0}} (00)
            edge   [bend right]            node [swap]{{1}} (10);
\end{tikzpicture}
    \caption{Example of a state-transition diagram that verifies that the proof of correctness of the algorithm proposed by Larsen is incomplete.}
    \label{state-transitiondiagram}
\end{figure}
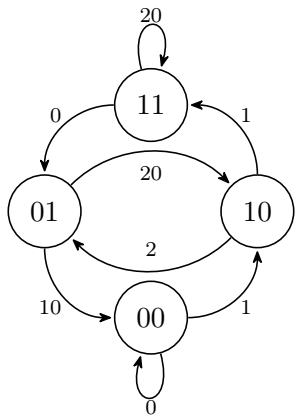

Before we introduce the algorithm proposed by Larsen, let us take into account that in this algorithm when a state is reached by a path it is stored in an array together with information on the type of the path (forward or backward) and the weight of the path. If there are many paths to a state, then the lowest weight is stored. The notation used is the same as in the new algorithm described in Section VI. For completeness we state the algorithm introduced by Larsen, see Algorithm \ref{alg:cap_larsen}.

\begin{algorithm}
\caption{ALGORITHM PRESENTED BY LARSEN }\label{alg:cap_larsen}
Let $W^\ast$ denote the current upper bound on the free distance. Then the algorithm for computing the free distance a $1/n$ convolutional code is the following.
\begin{algorithmic}
\State 1) Set $W^\ast$ as an upper bound on the free distance. Store $S=(1, 0 \ldots, 0)$, together with the weight $W_L$ of $S$ we get from the edge coming from $(0, \ldots, 0)$ and $T = F$ and $S = (0, 0 \ldots, 01)$, together with its weight $W_L$ we get from the edge going to $(0, \ldots, 0)$ and $T = B$ in the array.
\State 2) Search through the storage array for the lowest weight non$-D$ state $S$. It has weight $W_L$ and is of type $T$.
\State 3) If $2W \geq W^{\ast}$ or all states are $D$, go to 17).
\State 4) Set $E = 0$. Determine terminal state $S_E$ and weight $W_{E,L}$ of the $0$-extension of $S$. Go to 6).
\State 5) Set $E = 1$. Determine terminal state $S_E$ and weight $W_{E, L}$ of the $1$-extension of $S_m$.
\State 6)If $W_{E, L} > \frac{(W^{\ast} +n- 1)}{2}$, go to 15).
\State 7) Check through the array for $S_k = S_E$. If such an $S_k$ is found, go to 9).
\State 8) Find an unoccupied location in the array and store $S_E$,$W_{E, L}$, $T$. Go to 15).
\State 9) The type and weight of $S_k$  are $T_k$ and $W_{k, L}$. If ${T2}_k = {T2}$ go to 13).
\State 10) Set $W^\ast = \text{min}(W^\ast, W_{E, L} + W_{k, L})$.
\State 11) If $W_{E, L} \geq W_{k, L}$, go to 15).
\State 12) Set $W_{k, L} = W_{E, L}$ and ${T2}_{k} = {T2}_{m}$. Go to 15).
\State 13) If ${T1}_k = D$, go to 15).
\State 14) Set $W_{k, L} = \text{min} (W_{k, L}, W_{E, L})$.
\State 15) If $E = 0$, go to 5).
\State 16) Set ${T1} = D$. Go to 2).
\State 17) Return  $d_{free} = W^{\ast}$.
\end{algorithmic}
\end{algorithm}

Let us consider now for Example \ref{example} the steps of the algorithm. In step 1, we set, for example, $W^\ast = 100$ and we put $((1 0), 1, F)$ and $((01), 10, B)$ in the array.
In step 2, we set $S = (10), W_L = 1, T = F$.
The condition 3 is not satisfied so we go to step 4.
The $0$-extension is $S_0 = (01)$ and $W_{0, L} = 3$.
Condition 6 is not satisfied so we go to step 7.
Since $(01)$ is in the array, we go to step 9.
We set $S_k = (01), W_{k, L} = 10, T_k = B$.
Since $T2_k \neq T2$, we go to step 10 and set $W^\ast = 13$.
As $W_{0, L} < W_k$, we go to 12 and fix $W_{k, L} = 3$ and $T2_k = F$.
Therefore the array now contains $((1 0), 1, F)$ and $((0 1), 3, F)$.
Then we consider the $1$-extension of $S = (10)$ in the $F$ direction which is $S_1=(11)$.
So $((11),2,F)$ is stored in the array.
After that $S=(10)$ will be set dead and we go to step 2 again.
Here pick $S = (11), W_L = 2, T = F$.
Condition 3 is not satisfied, so we go to step 4. The $0$-extension of $S$ is $S_0 = (01)$ and $W_{0, L} = 2$.
Condition 6 is not valid and we find in 7 that $S_k = (0 1)$ is in the array.
In 9, we get $W_{k, L} = 3, T_k = F$ and go to step 13.
The condition in 13 is not satisfied and we go to 14.
In step 14, we set $W_{k, L} = 2$.
Afterwards we continue with the $1$-extension. Nothing changes in the array through the steps of the $1$-extension.
Afterwards we set $(11)$ dead.
Then in step 2 we go look for the lowest weight non-dead state in the array which is $S =(0 1)$ with $W_L=2$ and $T = F$.
In step 3 we continue.
We consider the $0$-extension of $S$ which is $S_0 = (00)$ with $W_{0, L}=12$.
Let us assume that the condition 6 is not satisfied (otherwise it becomes only easier). The state
$(0 0)$ is not in the array, so we store it with weight $12$.
Again in the step for the $1$-extension nothing changes in the array.
Then $(0 1)$ is set dead.
At this moment only $(0 0)$ is the non-dead state in the array.
Obviously $2 \cdot 12 = 24 > W^\ast = 13$ and we stop and return $13$.
This is not the distance of the shortest non-trivial path from $(0 0)$ to $(0 0)$. The correct answer would be $12$.
\end{example}

We will now describe the problem in the proof presented in \cite{larsen1973comments} and explain why the algorithm still works for convolutional codes.

Firstly, Larsen considers an arbitrary path in the state-transition diagram and uses a case analysis with two possibilities.
The first one is that all non-zero states in the path are dead and the second is that there are paths with some non-dead states.
The author claims that in the first case, the case with only dead states, there must be a part in the path where two adjacent states are of different $T2$ type.
Obviously in the example above this is not true.
In reality, it is simple to construct an example of a convolutional code where this is not the case, such as the code generated by $G(z) = (z^2 + 1, z^2)$.
Going through the algorithm we can observe, as previously, that the claim is incorrect for the code generated by $G(z)$. However now the algorithm returns the correct result, which is $3$. Let us elaborate a bit more. 

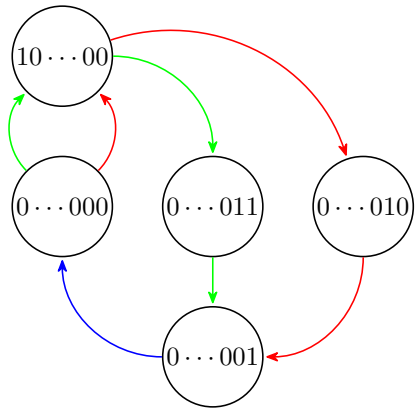
\begin{figure}[ht]
    \centering
    \begin{tikzpicture}[->,>={Stealth[round]},shorten >=1pt,
                    auto,node distance=2cm,on grid,semithick,
                    inner sep=1pt,bend angle=45,scale=0.6]
  \node[state,scale=1] (000)                    {$0\cdots 000$};
  \node[state,scale=1]         (100) [above =of 000] {$10\cdots 00$};
  \node[state,scale=1]         (011) [right=of 000] {$0\cdots011$};
  \node[state,scale=1]         (010) [right=of 011] {$0\cdots010$};
  \node[state,scale=1]         (001) [below=of 011] {$0\cdots001$};

  \path [every node/.style={font=\footnotesize}]
        (000)   edge    [bend right,red]                node{} (100)
                edge    [bend left, green]              node{} (100)
        (100)   edge    [bend left, scale=0.7, red]     node{} (010)
                edge    [bend left, green]              node{} (011)
        (011)   edge    [scale=0.7, green]              node{} (001)
        (010)   edge    [bend left, scale=0.7, red]     node{} (001)
        (001)   edge    [bend left, blue]               node{} (000);
\end{tikzpicture}
    \caption{Illustration on the idea to fix the proof in Larsen's paper for convolutional codes of rate $1/n$ over $\mathbb{F}_2$.}
    \label{state-transitiondiagram-proof-larsen}
\end{figure}

Let us assume that there are two previous states for the penultimate state $(0 \cdots 0 1)$, namely $(0 \cdots 0 1 1)$ and $(0 \cdots 0 1 0)$.
The crucial point for the counterexample is that there are $F$-paths from $(0 \cdots 0)$ through both $(0 \cdots 0 1 1)$ and $(0 \cdots 0  1 0)$ to $(0 \cdots 0 1)$ which have weight less than the weight of the edge from $(0 \cdots 0 1)$ to $(0 \cdots 0)$. More intuitively and considering Figure \ref{state-transitiondiagram-proof-larsen}, the weight of the green and red paths is smaller than the weight of the blue edge.

However, we claim that for a convolutional code it can never happen that a path from $(0 \cdots 0)$ through $(0 \cdots 0 1 1)$ to $(0 \cdots 0 1)$ has weight less than the edge from $(0 \cdots 0 1)$ to $(0 \cdots 0)$, i.e., we claim that the green path has weight at least the weight of the blue edge. To prove this, let us assume the opposite. Then at least one of the $n$ outputs (where $n$ is the code length) must give a $1$ on the blue edge and a $0$ on the green path.
Therefore it must be some linear combination of the registers with the following properties.
We write the linear combination in polynomial form $\sum_{i=0}^M a_i z^i$, where $a_i \in \mathbb{F}_2$ indicates which registers contribute to the output.
We need to have $a_M = 1$, otherwise the blue edge has a $0$ as output.
Furthermore, $a_{M-1} = 1$, otherwise the output for the transition from $(0 \cdots 0 1 1)$ to $(0 \cdots 0 1)$ would give output $1$ and this transition belongs to the green path.
Writing it as a polynomial we get $\sum_{i = 0}^{M-2} a_i z^i + z^{M-1} + z^{M}$.
But then any input will give weight at least $2$ which means that the green path %from $(0 \cdots 0)$ to $(0\cdots 0 1)$ via $(0 \cdots 0 1 1)$ 
has weight at least $1$ and so the claim follows.

In sum, we can conclude the correctness of the algorithm as follows.
Either the red path has at least the same weight as the blue edge. In this situation, Larsen's claim that in every path in which all non-zero states are dead there exist two adjacent states of different $T2$ type is true and the correctness remains as it is in the rest of Larsen's proof. If the red path has weight less than the blue edge, then in particular it has weight less than the green path as we have shown above.
This means that the adjustment of the upper bound through the red path and the blue edge that the algorithm does is valid, and there is no further need for an adjustment due to the green path.

Taking into account the return to the zero-state, which is important for $k > 1$, one can adapt the proof of Larsen also for the new algorithm proposed in Section VI.

\section*{Acknowledgment}

\small{This work is supported by the SNSF grant n. 212865, by CIDMA through FCT, \url{https://doi.org/10.54499/UIDB/04106/2020},
\url{https://doi.org/10.54499/UIDP/04106/2020} and by FCT grant UI/BD/151186/2021.}

\end{document}